\documentclass[12pt]{article}
\usepackage[english]{babel}
\usepackage[utf8]{inputenc}
\usepackage[T1]{fontenc}

\usepackage{amsmath}

\usepackage{amsfonts}
\def\tLambda{\tilde{\Lambda}}
\usepackage{url}
\usepackage[dvips]{graphicx}
\usepackage{calc}
\usepackage{subfigure}
\usepackage{multirow}

\def\mL{\mathcal{L}}
\def\bD{\mathbf{D}}
\def\mH{\mathcal{H}}
\def\mD{\mathcal{D}}

\def\bx{\mathbf{x}}

\newcommand{\pb}[1]{\left\{#1\right\}}

\def\tX{\tilde{X}}

\def\mG{\mathcal{G}}
\def\bG{\mathbf{G}}
\def\by{\mathbf{y}}

\def\bT{\mathbf{T}}

\def\mM{\mathcal{M}}

\begin{document}
	\begin{titlepage}
	\begin{center}
		{\Large{ \bf Reparametrization invariant action for Gravity with Dynamical
		Determinant of Metric}}
		
		\vspace{1em}  
		
		\vspace{1em} J. Kluso\v{n} 			
		\footnote{Email addresses:
			klu@physics.muni.cz (J.
			Kluso\v{n}) }\\
		\vspace{1em}
		\textit{Department of Theoretical Physics and
			Astrophysics, Faculty of Science,\\
			Masaryk University, Kotl\'a\v{r}sk\'a 2, 611 37, Brno, Czech Republic}
		
		\vskip 0.8cm
		
		%
		%
		%
		%
		%
		%
		
		\vskip 0.8cm
		
	\end{center}

\begin{abstract}
	We present manifestly reparametrization invariant action for theory of gravity 
	with dynamical determinant of metric. We show that it is similar to a reparametrization invariant action for unimodular gravity. We determine canonical
	form of the action and study structure of constraints.
\end{abstract}

\bigskip

\end{titlepage}

\newpage

\section{Introduction}\label{first}
Recently theories of gravities with dynamical determinant of metric
\cite{Alvarez:2006uu} were studied extensively in \cite{Kluson:2023bkg,Kluson:2023gnh}.
These theories can be considered as generalization of Weyl transverse gravity
\cite{Oda:2016psn} that has an important meaning in the formulation of unimodular gravity \cite{Buchmuller:1988wx,Henneaux:1989zc,Kuchar:1991xd}. In more details, Weyl transverse gravity
is invariant under local Weyl transformation whose gauge fixing leads to unimodular gravity, 
see for example 
\cite{Garcia-Moreno:2023zdk,Bengochea:2023dep,Alvarez:2023utn}. For that reason it is certainly 
very instructive to study theories which are more general than Weyl transverse gravity as theories with 
dynamical determinant of metric are. The preliminary step in their investigation was performed in 
\cite{Kluson:2023bkg,Kluson:2023gnh} where we have shown that in case of general form of gravity
with dynamical determinant of metric the theory seems to be non-consistent in the sense
that Hamiltonian constraint is the second class constraint. The similar situation occurs in case
of non-projectable Ho\v{r}ava-Lifshitz gravity \cite{Horava:2009uw}
as was shown in \cite{Blas:2009yd,Li:2009bg,Henneaux:2009zb} with possible solution found in \cite{Blas:2009qj,Kluson:2010nf}. We speculated in \cite{Kluson:2023bkg} that possible origin 
of this non-consistency is hidden in manifest non-covariance of the theory with dynamical determinant of metric. In fact, the similar situation occurs in case of unimodular gravity as was shown very recently in 
\cite{Bengochea:2023dep}. It was argued there  that the proper formulation of unimodular gravity needs some fixed background
volume form. Then  we studied Weyl transverse gravity with fixed background field and we were able to show
that this theory is consistent from the point of view of canonical formalism 
\cite{Kluson:2023ujn} and that the gauge fixing of local Weyl symmetry leads to the unimodular gravity. 

Even if the presence of the background volume form makes the theory invariant under diffeomorphism transformations it is not completely clear how to deal with such fields in canonical formulation. For example, with the presence of the background fields we should carefully distinguish between active and passive diffeomorphism transformations, see very nice recent discussion in 
\cite{Garcia-Moreno:2023zdk}.We also mean that  the presence of privileged background structure seems to be against physical intuition where we expect that the action should contain all physical fields in nature and we should not a priori select some of them to be fixed. 
For that reason we mean that much more intuitive treatment of theories of gravity with dynamical metric is based on idea of reparametrization that was introduced long time ago by K. Kucha\v{r} in case of unimodular gravity \cite{Kuchar:1991xd}. In this article we implement this approach to the case of theories of gravity with dynamical determinant of metric. 

The basic idea is simple. Following \cite{Kuchar:1991xd} we start with the action with dynamical determinant of metric written in coordinates, that we call generalized unimodular coordinates $X^A$, where this action lacks manifest diffeomorphism invariance. The invariance of theory is restored by parametrization where the generalized unimodular coordinates are replaced with arbitrary coordinates $x^\mu$  while unimodular coordinates become scalar fields $X^A(x^\mu)$. Then we define parametrized action which is invariant under general diffeomorphism transformation. Such an action contains a term which is proportional to the Jacobian of transformation from new variables to generalized unimodular ones. 

Having such a manifestly diffeomorphism invariant action we can proceed to the canonical formulation of 
theories of gravity with dynamical determinant of metric. It turns out that this action has a form of 
gravity coupled to the scalar field together with an action for $X^A$. Following 
\cite{Kuchar:1991xd} we perform two possible rearrangements of this action. In the fist one we obtain 
formulation of theory in the form of Henneaux-Teitelboim covariant formulation of unimodular gravity 
\cite{Henneaux:1989zc}. Then careful canonical analysis of this theory leads to the result that 
theory of gravity with dynamical determinant of metric is equivalent to unimodular theory of gravity coupled to scalar field. This conclusion is also confirmed by the second rearrangement of the action for scalar field $X^A$ where we follow \cite{Kuchar:1991xd}. We again find canonical formulation of theory and we find that it is equivalent to the unimodular gravity coupled with scalar field. 

This paper is organized as follows. In the next section (\ref{second}) we introduce general action of gravity with dynamical determinant of metric. Then we rewrite it in manifest diffeomorphism invariant form
and perform its canonical formulation.  In section (\ref{third}) we analyse this theory with second possible rearrangement of the action for scalar fields $X^A$. In conclusion (\ref{fourth}) we outline our results. Finally in Appendix A we present explicit calculations of  Poisson bracket between Hamiltonian functions where we show that it vanishes on the constraint surface.

\section{Action with Dynamical  Determinant of Metric}
\label{second}
In this section we review basic facts about gravity with dynamical determinant of metric. Generally this action has the form \cite{Alvarez:2006uu,Lopez-Villarejo:2010uib}
\begin{eqnarray}\label{Sgen}
&&S=\int d^{4}X\mL \ , \quad 
\mL=\frac{1}{\kappa} F(\sqrt{-g})\sqrt{-g}[R(g_{AB})+\nonumber \\
&&+G(\sqrt{-g})\partial_A \sqrt{-g}g^{AB}
\partial_B \sqrt{-g}] \ , \quad \nonumber \\
&& g\equiv \det g_{AB} \ , \quad \kappa=16\pi \ , 
\end{eqnarray}
where $F$ and $G$ are arbitrary functions of $\sqrt{-g}$. Note that we work in 
$4$-dimensional space-time with metric signature $(-,+,+,+)$ and $A,B=0,1,2,3$.
It is straightforward to extend this analysis to general dimensions however making notation simple it is sufficient to work in four dimensions only. 
 
Following  
\cite{Kuchar:1991xd} we call $X^A=(T,Z^i),i,j=1,2,3$ and we denote them as generalized unimodular coordinates. Let us introduce new scalar field $\phi$ and  rewrite the action into the form
\begin{eqnarray}\label{Sgenr}
	&&S=\frac{1}{\kappa}\int d^{4}X
[ F(\phi)\sqrt{-g}[R(g_{AB})+G(\phi)\partial_A  \phi g^{AB}
	\partial_B\phi ]+\Lambda(\phi-\sqrt{-g})] \ . \quad \nonumber \\
\end{eqnarray}
It is clear from (\ref{Sgenr}) that  the action is invariant under diffeomorphism transformation 
up to the term $\Lambda\phi$ since  $\Lambda$ and  $\phi$ are scalars. Following 
\cite{Kuchar:1991xd} we can restore full diffeomorphism
invariance by parametrization. The privileged generalized unimodular variables $X^A$ are replaced
with arbitrary coordinates $x^\mu$ and are promoted to the field variables $X^A=X^A(x^\mu)$. Then as in 
\cite{Kuchar:1991xd}  we introduce Jacobian of transformation from $x^\mu\rightarrow X^A$ 
\begin{equation}
\tX=\frac{1}{4!}
\epsilon_{ABCD}X^A_\alpha X^B_\beta X^C_\gamma X^D_\delta \epsilon^{\alpha\beta\gamma\delta}  ,  \quad X^A_\alpha\equiv \frac{\partial X^A}{\partial x^\alpha} \ .
\end{equation}
Then the  reparametrization invariant action has the form 
\begin{eqnarray}\label{Sgenrep}
&&S^{rep}=\frac{1}{\kappa}\int d^{4}x
[ F(\phi)\sqrt{-g}[R(g_{\mu\nu})+G(\phi)\partial_\mu  \phi g^{\mu\nu}
\partial_\nu\phi ]+\Lambda(\phi \tX-\sqrt{-g})] \ . \quad \nonumber \\
\end{eqnarray}
Following 
\cite{Kuchar:1991xd} we bring the action into canonical form. Let us start with the gravitational part that has the form 
\begin{eqnarray}\label{Srepgr}
&&S^{rep,GR}=\frac{1}{\kappa}\int d^{4}x
 F(\phi)\sqrt{-g}[R(g_{\mu\nu})+G(\phi)\partial_\mu  \phi g^{\mu\nu}
\partial_\nu\phi -\frac{\Lambda}{F(\phi)}] \ . 
\nonumber \\ 
\end{eqnarray}
In order to find canonical form of action we implement 
 well known $3+1$
formalism that is the fundamental ingredient of the Hamiltonian
formulation of any theory of gravity \footnote{For recent review, see
	\cite{Gourgoulhon:2007ue}.}. We consider $4-$dimensional manifold
$\mathcal{M}$ with the coordinates $x^\mu \ , \mu=0,\dots,3$ and
where $x^\mu=(t,\bx) \ , \bx=(x^1,x^2,x^{3})$. Let  $ \mathcal{M}$ is
foliated by a family of space-like surfaces $\Sigma_t$ defined by
$t=x^0=\mathrm{const}$. Let $h_{ij}$ denotes the metric on $\Sigma_t$
with inverse $h^{ij}$ so that $h_{ij}h^{jk}= \delta_i^k$. We further
introduce the operator $\nabla_i$ that is covariant derivative
defined with the metric $h_{ij}$.
Let us define $n^\mu$ as 
future-pointing unit normal vector  to the surface
$\Sigma_t$. In ADM variables we have $n^0=\sqrt{-g^{00}},
n^i=-g^{0i}/\sqrt{-g^{ 00}}$.
We also define  the lapse
function $N=1/\sqrt{-g^{00}}$ and the shift function
$N^i=-g^{0i}/g^{00}$. In terms of these variables we
write the components of the metric $g_{\mu\nu}$ as
\begin{eqnarray}
&&g_{00}=-N^2+N_i h^{ij}N_j \ , \quad g_{0i}=N_i \ , \quad
g_{ij}=h_{ij} \ ,
\nonumber \\
&&g^{00}=-\frac{1}{N^2} \ , \quad g^{0i}=\frac{N^i}{N^2} \
, \quad g^{ij}=h^{ij}-\frac{N^i N^j}{N^2} \ 
\nonumber \\
\end{eqnarray}
and hence $g=-N^2\det h$. We further have following decomposition of $R$ in the form
\begin{eqnarray}
&&R=K^{ij}K_{ij}-K^2+r+2\tilde{\nabla}_\mu[\tilde{n}^\mu K]
-\frac{2}{N}\nabla_i\nabla^iN \ , 
\nonumber \\
&&K_{ij}=\frac{1}{2N}(\partial_t h_{ij}-\nabla_i N_j-\nabla_j N_i) \ , 
\quad n^0=\sqrt{-g^{00}} \ , \quad 
n^i=-\frac{g^{0i}}{\sqrt{-g^{00}}} \ , \nonumber \\
\end{eqnarray}
and where $r$ is scalar curvature defined with $h_{ij}$ and where $\tilde{\nabla}_\mu$ is covariant derivative compatible with $g_{\mu\nu}$  while 
$\nabla_i$ is covariant derivative compatible with the metric $h_{ij}$. Note that we can also write
\begin{eqnarray}
&&	\tilde{\nabla}_\mu [n^\mu K]=\frac{1}{\sqrt{-g}}\partial_\mu[\sqrt{-g}n^\mu K]
\ , \nonumber \\
&&\partial_\mu \phi g^{\mu\nu}\partial_\nu \phi=
-\nabla_n\phi^2+
h^{ij}\partial_i \phi\partial_j \phi \ , \quad 
\nabla_n\phi=\frac{1}{N^2}(\partial_0 \phi-N^i\partial_i \phi) \ .  \nonumber \\
\end{eqnarray}
With the help of this notation and using integration by parts we can rewrite the action 
 $S^{rep,GR}$ into the form  
\begin{eqnarray}\label{SKK}
&&S^{rep,GR}
=\frac{1}{\kappa}
\int dt d^3\bx (FN\sqrt{h}[K_{ij}\mG^{ijkl}K_{kl}+r]
-2F'N\sqrt{h}\nabla_n\phi K-FGN\sqrt{h}\nabla_n\phi^2-\nonumber \\
&&-2N\partial_i[F'\sqrt{h}h^{ij}\partial_j \phi ]+
FGN\sqrt{h}h^{ij}\partial_i\phi\partial_j\phi-\Lambda N\sqrt{h})
\nonumber \\
\end{eqnarray}
or equivalently  
\begin{eqnarray}\label{SGReq}
&&S^{rep,GR}
=\frac{1}{\kappa}
\int dt d^3\bx (FN\sqrt{h}K_{ij}\mM^{ijkl}K_{kl}+
N\sqrt{h}r-FGN\sqrt{h}\tX^2-
\nonumber \\
&&-2N\partial_i[F'\sqrt{h}h^{ij}\partial_j \phi ]+
FGN\sqrt{h}h^{ij}\partial_i\phi\partial_j\phi-\Lambda N\sqrt{h}) \ , 
\nonumber \\
\end{eqnarray}
where
\begin{eqnarray}
&&\tX=\nabla_n\phi+\frac{F'}{FG}K \ , \quad F'\equiv \frac{dF}{d\phi} \  , \nonumber \\
&&\mM^{ijkl}=\mG^{ijkl}+\frac{F'^2}{F^2G}h^{ij}h^{kl}=
\frac{1}{2}(h^{ik}h^{jl}+h^{il}h^{jk})-h^{ij}h^{kl}(1-\frac{F'^2}{F^2G}) \ .
\nonumber \\
\end{eqnarray}
Now from (\ref{SGReq}) we obtain  momenta conjugate to $h_{ij}$ and $\phi$
\begin{eqnarray}\label{momenta}
&&\pi^{ij}=\frac{\partial \mL^{rep,GR}}{\partial(\partial_t h_{ij})}=
\frac{1}{\kappa}F\sqrt{h}(\mM
^{ijkl}K_{kl}-\frac{F'}{F}h^{ij}\tX) \ ,  \nonumber \\
&&p_\phi=\frac{\partial \mL^{rep,GR}}{\partial(\partial_t \phi)}=
-\frac{2}{\kappa}FG\sqrt{h}\tX  \nonumber \\
\end{eqnarray}	
so that $H^{rep,GR}$ is equal to 
\begin{eqnarray}\label{HrepGR}
&&H^{rep,GR}=\int d^3\bx (\pi^{ij}\partial_t h_{ij}+p_\phi \partial_t\phi-\mL^{rep,GR})=
\nonumber \\
&&=\frac{1}{\kappa}\int d^3\bx (N\sqrt{h}FK_{ij}\mM^{ijkl}K_{kl}-
FGN\sqrt{h}\tX^2
-FN\sqrt{h}r-\nonumber \\
&&-FGN\sqrt{h}h^{ij}\partial_i\phi\partial_j\phi+
2N\partial_i[F'\sqrt{h}h^{ij}\partial_j\phi]+\Lambda N\sqrt{h}+N^i\mH_i) \ , 
\nonumber \\
\end{eqnarray}
where
\begin{eqnarray}
	\mH_i=p_\phi \partial_i\phi-2 h_{ik}\nabla_j\pi^{jk} \ . 
	\nonumber \\
\end{eqnarray}
Clearly we should express Hamiltonian as functions of canonical variables 
$(\pi^{ij},h_{ij}),(\phi,p_\phi)$. To do this we firstly define $\Pi^{ij}$ as 
\begin{equation}\label{defPi}
\Pi^{ij}\equiv \pi^{ij}-\frac{1}{2}\frac{F'}{FG}p_\phi h^{ij} \ . 
\end{equation}
Then from (\ref{momenta}) we get 
\begin{eqnarray}\label{Piij}
\Pi^{ij}=
\frac{1}{\kappa}F\sqrt{h}\mM^{ijkl}K_{kl} \ . 
\nonumber \\
\end{eqnarray}
Next step depends on the fact whether the matrix $\mM^{ijkl}$ is invertible or not. 
Let us for the time being presume that it is so that there exists inverse matrix
$\mM_{ijkl}$ that obeys
\begin{equation}\label{defMijkl}
\mM_{ijkl}\mM^{klmn}=\frac{1}{2}(\delta_i^m\delta_j^n+\delta_i^n\delta_j^m) \ . 
\end{equation}
Then from (\ref{Piij}) we express $K_{ij}$ as functions of canonical variables 
\begin{eqnarray}\label{Kij}
K_{ij}=\frac{\kappa}{F\sqrt{h}}\mM_{ijkl}\Pi^{kl} \ ,\nonumber \\
\end{eqnarray}
where the matrix $\mM_{ijkl}$ is equal to 
\begin{equation}
		\mM_{ijkl}=\frac{1}{2}(h_{ik}h_{jl}+h_{il}h_{jk})-
\left(\frac{1-\frac{F'^2}{F^2G}}{2-3\frac{F'^2}{F^2G}}	\right)
		h_{ij}h_{kl} \ . 
	\end{equation}
Note that this result is well defined on condition $2-3\frac{F'^2}{F^2G}\neq 0$. 
Finally using (\ref{Kij}) in (\ref{HrepGR}) we get final form of the Hamiltonian
\begin{eqnarray}
&&H^{rep,GR}=\int d^3\bx (N\mH^{rep,GR}+N^i\mH^{rep,GR}_i) \ , 
\nonumber \\
&&\mH^{rep,GR}=
\frac{\kappa}{\sqrt{h}F}\Pi^{ij}\mM_{ijkl}
\Pi^{kl}-\frac{\kappa}{4GF\sqrt{h}}p_\phi^2
-\frac{1}{\kappa}F\sqrt{h}r-\nonumber \\
&&-\frac{1}{\kappa}FG\sqrt{h}h^{ij}\partial_i\phi\partial_j\phi+
\frac{2}{\kappa}N\partial_i[F'\sqrt{h}h^{ij}\partial_j\phi]+\frac{1}{\kappa}\Lambda N\sqrt{h}\equiv 
\nonumber \\
&&\equiv \mH_\tau+\frac{1}{\kappa}\Lambda \sqrt{h} \ . \nonumber \\
\end{eqnarray}
\subsection{The Case of Singular Matrix $\mM^{ijkl}$}
In this section we will discuss the case when the matrix $\mM^{ijkl}$ is singular.
As in non-singular case we  find 
%
\begin{equation}\label{PiMsin}
\Pi^{ij}=\frac{1}{\kappa}F\sqrt{h}\mM^{ijkl}K_{kl} \ ,
\end{equation}
where $\Pi^{ij}$ is defined in (\ref{defPi}).
Let us now presume that $\mM^{ijkl}$ cannot be inverted and that $h_{ij}$ is zero eigenvector of $\mM^{ijkl}$
\begin{equation}
\mM^{ijkl}h_{kl}=h^{ij}\left(-2+3\frac{F'^2}{F^2G}\right)=0
\end{equation}
so that
\begin{equation}
\frac{F'^2}{F^2G}=\frac{2}{3} \ . 
\end{equation}
Using this explicit value we find 
\begin{equation}\label{Mproj}
\mM^{ijkl}=\frac{1}{2}(h^{ik}h^{jl}+h^{il}h^{jk})-
\frac{1}{3}h^{ij}h^{kl} \ 
\end{equation}
and hence
\begin{eqnarray}
\mM^{ijkl}h_{km}h_{ln}\mM^{mnpr}=\mM^{ijpr} \ . 
\end{eqnarray}
Note that using this relation we obtain from (\ref{PiMsin}) that there is a primary constraint  $\mD$ defined as
\begin{equation}
\mD\equiv h_{ij}\Pi^{ij}=\pi^{ij}h_{ij}-\frac{F}{F'}p_\phi\approx 0 \ . 
\end{equation}
Further, using (\ref{Mproj}) we can find $\mH^{rep,GR}_D$ in the form 
\begin{eqnarray}\label{repGRD}
&&\mH^{rep,GR}_D
=\frac{\kappa }{F\sqrt{h}}\Pi^{ij}h_{ik}h_{jl}\Pi^{kl}-
\frac{\kappa}{4F\sqrt{h}}\frac{2}{3}\frac{F^2}{F'^2}p_\phi^2
-\frac{1}{
	\kappa}F\sqrt{h}r-\nonumber \\
&&-\frac{1}{\kappa}F\frac{3}{2}\frac{F'^2}{F^2}\sqrt{h}h^{ij}\partial_i\phi\partial_j\phi+
\frac{2}{\kappa}\partial_i[F'\sqrt{h}h^{ij}\partial_j\phi]
+\frac{1}{\kappa}\Lambda\sqrt{h} \ . 
\nonumber \\
\end{eqnarray}
Let us now discuss properties of the constraint $\mD\approx 0$ in more details. It is instructive to introduce its smeared form  $\bD(\Omega)$ as
\begin{equation}
\bD(\Omega)=\int d^3\bx \Omega \mD \ . 
\end{equation}
Then for further purposes we calculate Poisson bracket between 
$\bD(\Omega)$ and $\mH^{rep,GR}_D$ given in (\ref{repGRD}). Using canonical Poisson brackets 
\begin{equation}\label{canPB}
\pb{h_{ij}(\bx),\pi^{kl}(\by)}=\frac{1}{2}(\delta_i^k\delta_j^l+\delta_j^k
\delta_i^l)\delta(\bx-\by) \ , \quad 
\pb{\phi(\bx),p_\phi(\by)}=\delta(\bx-\by)
\end{equation}
we obtain
\begin{eqnarray}
&&\pb{\bD(\Omega),h_{ij}}=-h_{ij}\Omega \ , \quad 
\pb{\bD(\Omega),\pi^{ij}}=\Omega \pi^{ij} \ , \nonumber \\
&&\pb{\bD(\Omega),\phi}=\Omega \frac{F}{F'} \ , \quad \pb{\bD(\Omega),\frac{F}{F'}p_\phi}=0 \ , 
\nonumber \\ 
&&\pb{\bD(\Omega),\Pi^{ij}}=\Pi^{ij}\Omega \ , \quad 
\pb{\bD(\Omega),F(\phi)}=F\Omega \ , 
\nonumber \\
&&\pb{\bD(\Omega),h^{ij}}=\Omega h^{ij} \ ,  \quad 
\pb{\bD(\Omega),\sqrt{h}}=-\frac{3}{2}\sqrt{h}\Omega \ . \nonumber \\
\end{eqnarray}
Now we are ready to proceed to the calculation of Poisson bracket between $\bD$ and $\mH^{rep,GR}_D$ given in (\ref{repGRD}).
First of all we have
\begin{eqnarray}
&&\pb{\bD(\Omega),\frac{\kappa}{F\sqrt{h}}\Pi^{ij}h_{ik}h_{jl}\Pi^{kl}}=
\frac{1}{2}\Omega \frac{\kappa}{F\sqrt{h}}\Pi^{ij}h_{ik}h_{jl}\Pi^{kl} \ , \nonumber \\
&&\pb{\bD(\Omega),\frac{\kappa}{4F\sqrt{h}}\frac{2}{3}\frac{F^2}{F'^2}p_\phi^2}=
\frac{1}{2}\Omega \frac{\kappa}{4F\sqrt{h}}\frac{2}{3}\frac{F^2}{F'^2}p_\phi^2 \ , \nonumber \\
&&\pb{\bD(\Omega),-\frac{1}{\kappa}F\sqrt{h}r}=
-\frac{1}{2}\Omega \frac{1}{\kappa}F\sqrt{h}r-\frac{2}{\kappa}F\sqrt{h}\nabla_k\nabla^k\Omega
\nonumber \\
\end{eqnarray}
using
\begin{equation}
\frac{\delta r(\bx)}{\delta h_{ij}(\by)}=
-r^{ij}(\bx)\delta(\bx-\by)+\nabla^i\nabla^j \delta(\bx-\by)
-h^{ij}\nabla_k\nabla^k\delta(\bx-\by) \ . 
\end{equation}
Further we have
\begin{eqnarray}
\pb{\bD(\Omega),-\frac{1}{\kappa}F\frac{3}{2}\frac{F'^2}{F^2}
	\sqrt{h}h^{ij}\partial_i\phi\partial_j\phi}
=-\frac{1}{2\kappa}
F\frac{3}{2}\frac{F'^2}{F^2}\sqrt{h}h^{ij}\partial_i\phi\partial_j\phi
-\frac{3}{\kappa}F'\sqrt{h}\partial_i\Omega h^{ij}\partial_j\phi\nonumber \\
\end{eqnarray}	
and 
\begin{eqnarray}
&&\pb{\bD(\Omega),\frac{2}{\kappa}\partial_i[F'\sqrt{h}h^{ij}\partial_j\phi]}=
\nonumber \\
&&=\frac{1}{2}\frac{2}{\kappa}\Omega \partial_i[F'\sqrt{h}h^{ij}\partial_j\phi]
+\frac{3}{\kappa}F'\sqrt{h}\partial_i\Omega h^{ij}\partial_j\phi+
\frac{2}{\kappa}F\partial_i[\sqrt{h}h^{ij}\partial_j\Omega] \ .  \nonumber \\
\end{eqnarray}
For further purposes we perform rescaling $\Lambda=\frac{\tLambda}{\phi}$ 
and  calculate
\begin{equation}
\pb{\bD(\Omega),\frac{\tLambda}{\phi}\sqrt{h}}=-\frac{\tLambda}{\phi}
\sqrt{h}(\frac{3}{2}+\frac{1}{\phi}\frac{F}{F'}) \ . 
\end{equation}
Then collecting all these terms together we obtain 
\begin{eqnarray}\label{pbbDmHrep}
\pb{\bD(\Omega),\mH^{rep,GR}_D}=\frac{1}{2}\Omega\mH_\tau
-\frac{\tLambda}{\phi}
\sqrt{h}(\frac{3}{2}+\frac{1}{\phi}\frac{F}{F'})\neq \frac{1}{2}\Omega \mH^{rep,GR}_D \ . 
\end{eqnarray}
We see that this Poisson bracket is proportional  to $\mH^{rep,GR}_D$ on condition when
\begin{equation}\label{cond}
\frac{3}{2}+\frac{1}{\phi}\frac{F}{F'}=-\frac{1}{2} \ . 
\end{equation}
This can be interpreted as differential equation for $F$ that 
has simple solution
\begin{equation}
 F=\phi^{-\frac{1}{2}} \ . 
\end{equation}
It turns out that this is precisely the case of Weyl transverse gravity 
\cite{Oda:2016psn} that was also recently studied in \cite{Kluson:2023ujn}.
 In fact, we will see in the next
section that the condition (\ref{cond}) is crucial for the preservation 
of the constraint $\mD\approx 0$ during the time evolution of system. 
\subsection{The Reparametrization Term-The First Rearrangement}
As the final step we consider the reparametrization term
\begin{eqnarray}\label{S2}
&&S^{rep,\tX}=\frac{1}{\kappa}\int d^4x \Lambda \phi \tX \ , \nonumber \\
&&\tX=\frac{1}{4!}\epsilon_{ABCD}X^A_\alpha X^B_\beta X^C_\gamma X^D_\delta
\epsilon^{\alpha\beta\gamma\delta} \ , \quad X^A_\alpha=\frac{\partial X^A}{\partial x^\alpha} \ . 
\end{eqnarray}
There are two possibilities how to deal with $\tX$ \cite{Kuchar:1991xd}. In this section we rearrange  $\tX$ as
\begin{eqnarray}
&&\tX=\frac{1}{3!}\epsilon_{0IJK}\partial_\alpha T Z^I_\beta Z^J_\gamma
Z^K_\delta \epsilon^{\alpha\beta\gamma\delta}=
\nonumber \\
&&=\frac{1}{3!}\epsilon_{IJK}\partial_\alpha [T Z^I_\beta Z^J_\gamma
Z^K_\delta \epsilon^{\alpha\beta\gamma\delta}]
-\frac{1}{3!}\epsilon_{IJK}T\partial_\alpha ( Z^I_\beta Z^J_\gamma
Z^K_\delta \epsilon^{\alpha\beta\gamma\delta})=\nonumber \\
&&=
\partial_\alpha \tau^\alpha \ , \quad 
\tau^\alpha=\frac{1}{3!}\epsilon_{IJK}T Z^I_\beta Z^J_\gamma
Z^K_\delta \epsilon^{\alpha\beta\gamma\delta} \ ,
\nonumber \\
\end{eqnarray}
where $X^A=(T,Z^K), I,J,K=1,2,3$ and where we used the fact that $\partial_\alpha
Z^K_\beta=\partial_\alpha\partial_\beta Z^K$ is symmetric in $\alpha,\beta$ indices while $\epsilon^{\alpha\beta\gamma\delta}$ is antisymmetric when we exchange $\alpha$ and $\beta$. Note that we can equivalently use $\tau^\alpha$ as independent variables
instead of $X^A$ when we pass to the Hamiltonian formalism.  However looking on the form
of the action  (\ref{S2}) it is convenient to perform integration by part in case of $
\tau^\alpha$ and also replace $\Lambda=-\frac{\mu}{\phi}$ so that we get
\begin{eqnarray}
&&S^{rep,\tX}=-\frac{1}{\kappa}\int d^4x \partial_\alpha \mu \tau^\alpha=
\frac{1}{\kappa}\int d^4x [\partial_0 \mu \tau^0+\partial_i\mu \tau^i] \ .
\nonumber \\
\end{eqnarray}
We see that $S^{rep,\tX}$  has already canonical form and  it is natural to identify $\tau^0$ as momentum conjugate to 
 $\mu$ while $\tau^i$ are Lagrange multipliers that ensure that $\mu$ depends on time only. In the same
 way we can interpret $N$ and $N^i$ as Lagrange multipliers 
corresponding to the constraints $\mH^{rep,GR},\mH^{GR}_i$ so that the true physical phase space is 
\begin{equation}
 (h_{ij},\pi^{ij}) \ , \quad (\mu,\tau^0)	, \quad  (\phi,p_\phi) \ . 
\end{equation}
Then the canonical form of the action is 
\begin{eqnarray}
&&	S=\int dt d^3\bx (\partial_t h_{ij}\pi^{ij}+\partial_t\phi p_\phi+\tau^0\partial_t \mu-
	N \mH^{rep,GR}-N^i\mH_i-\tau^i\partial_i\mu) =\nonumber \\
&&=\int dt d^3\bx (\partial_t h_{ij}\pi^{ij}+\partial_t\phi p_\phi+\tau^0\partial_t \mu-
H_T)
	\nonumber \\
\end{eqnarray}
so that we have following set of constraints
\begin{equation}
	\mH^{rep,GR}\approx 0 \ , \quad \mH_i\approx 0 \ , \quad \partial_i\mu \approx 0 \ . 
\end{equation}
In the next section we will study stability of these constraints.
%
%

 \subsection{Stability of the Constraints}
The constraint is stable when it is preserved during the time evolution of the system. Since
time evolution of any function on the phase space is given by following equation
\begin{equation}\label{defF}
	\partial_t F=\pb{F,H_T}
\end{equation}
we see that the constraint is preserved during time evolution on condition when the Poisson 
bracket between this constraint and Hamiltonian vanishes on the constraint surface where the constraint
surface corresponds to the surface where all constraints vanish. 
As we argued above the Hamiltonian is given as sum of the constraints $\mH^{rep,GR}\approx 0, \mH_i\approx 0$ and $\partial_i\mu\approx 0$. Then it is clear that the  constraints are preserved during the time evolution when their Poisson brackets with all  constraints
vanish on the constraint surface.

Let us start with the constraint $\mH_i\approx 0$ where it is convenient to introduce
their smeared form 
\begin{equation}
\bT_S(\xi^i)=\int d^3\bx \xi^i\mH_i \ . 
\end{equation}
Then with the help of following Poisson brackets 
\begin{eqnarray}\label{bTSh}
&&	\pb{\bT_S(\xi^i),h_{ij}}=
-\xi^m\partial_m h_{ij}-\partial_i \xi^m h_{mj}-
h_{im}\partial_j\xi^m \ , \nonumber \\
&&	\pb{\bT_S(\xi^i),\pi^{ij}}=-\partial_m(\xi^m\pi^{ij})+
\partial_m \xi^i\pi^{mj}+\pi^{im}\partial_m \xi^j \ , \nonumber \\
&&	\pb{\bT_S(\xi^i),\phi}=-\xi^m\partial_m\phi \ , \nonumber \\
&&	\pb{\bT_S(\xi^i),p_\phi}=-\partial_i\xi^i p_\phi-\xi^m\partial_m p_\phi \ , 
\nonumber \\
&&\pb{\bT_S(\xi^i),\sqrt{h}}=-\xi^m\partial_m\sqrt{h}-\partial_m\xi^m\sqrt{h}
\nonumber \\
\end{eqnarray}
we get
\begin{equation}
\pb{\bT_S(\xi^i),\bT_S(\zeta^j)}=
\bT_S(\xi^m\partial_m\zeta^n-\zeta^m\partial_m \xi^n)
\end{equation}
that vanishes on the constraint surface $\mH_i\approx 0$. Further, it is also clear that 
\begin{equation}
\pb{\bT_S(\xi^i),\partial_j\mu(\bx)}\approx 0  \ . 
\end{equation}
Finally  we calculate Poisson bracket between $\bT_S(\xi^i)$ and $\mH^{rep,GR}\approx 0$. It is convenient to write $\mH^{rep,GR}$ as
\begin{equation}
\mH^{rep,GR}\equiv \mH_\tau-\frac{\mu}{\phi}\sqrt{h} \ . 
\end{equation}
Then we obtain 
\begin{eqnarray}
&&\pb{\bT_S(\xi^i),\mH^{rep,GR}}=-\xi^m\partial_m\mH_\tau-\partial_m\xi^m\mH_\tau
+\mu\xi^m\partial_m[\frac{\sqrt{h}}{\phi}]+\partial_m\xi^m\frac{\mu}{\phi}\sqrt{h}=
\nonumber \\
&&=-\xi^m\partial_m\mH^{rep,GR}-\partial_m\xi^m\mH^{rep,GR}-
\xi^m\partial_m\mu \frac{\sqrt{h}}{\phi}\approx 0 \  \nonumber \\
\end{eqnarray}
that again vanishes on the constraint surface $\mH_i\approx 0 \ , \partial_m\mu\approx 0$. 
In other words we checked that $\mH_i\approx 0$ and $\partial_i\mu$ are first class constraints. 

Now we come to the most difficult part with is proof that $\mH^{rep,GR}$
is first class constraint too.  It is again convenient to introduce smeared form of constraints $\mH_\tau$ 
\begin{equation}
\bT_T^{GR}(\xi)=\int d^3\bx \xi(\bx) \mH_\tau(\bx) \ , \quad 
\bT_T^{GR}(\zeta)=\int d^3\bx \zeta(\bx) \mH_\tau(\bx) \ , 
\end{equation}
where 
\begin{eqnarray}
&&\mH_\tau=
\frac{\kappa}{\sqrt{h}F}\Pi^{ij}\mM_{ijkl}
\Pi^{kl}-\frac{\kappa}{4GF\sqrt{h}}p_\phi^2
-\frac{1}{\kappa}F\sqrt{h}r-\nonumber \\
&&-\frac{1}{\kappa}FG\sqrt{h}h^{ij}\partial_i\phi\partial_j\phi+
\frac{2}{\kappa}\partial_i[F'\sqrt{h}h^{ij}\partial_j\phi] \ , 
\nonumber \\
\end{eqnarray}
where we now consider  the case when the matrix $\mM^{ijkl}$ is non-singular. 
The calculations with singular case are completely the same so that we will not explicitly write here. The more detailed calculations will be presented in Appendix 
with the result
\begin{equation}
	\pb{\bT^{GR}_T(\xi),\bT^{GR}_T(\zeta)}=
	\bT_S((\xi\partial_m\zeta-\zeta\partial_m\xi)h^{mn}) \ . 
 \end{equation}
 Using this result it is easy to see that $\mH^{rep,GR}\approx 0$ is the first class constraints together with $\mH_i\approx 0$ and $\partial_i\mu\approx 0$. Note that 
 the last condition implies that $\mu$ as dynamical variable depends on time only. Then however using the fact that Hamiltonian does not depend on $\tau^0$ explicitly we get that $\mu$ is constant during the time evolution of system. 
 
 Let us now consider the second case when the matrix $\mM^{ijkl}$ is singular. As we argued before this leads to an emergence of the primary constraint $\mD\approx 0$ that has non-zero Poisson bracket with $\mH_\tau\approx 0$ given in 
(\ref{pbbDmHrep}). We see that it is natural to demand that this Poisson bracket 
vanishes on the constraint surface so that $\mD\approx 0$ and $\mH^{rep,GR}\approx 0$ are first class constraints. As we argued in previous section this  case corresponds to Weyl transverse gravity. Note that the opposite case would imply that $\mD\approx 0$ and $\mH^{rep,GR}\approx 0$ are second class constraints which would imply inconsistency of theory with lack of the
generator of time reparametrization as the first class constraint. 

In the next section we consider second possible form of reparametrization term and analyse  structure of theory.
\section{Second Rearrangement}\label{third}
In this section we consider the second rearrangement of the action $S^{rep,\tX}$ that was introduced in \cite{Kuchar:1991xd}. Recall that  $S^{rep,\tX}$ has the form  
\begin{equation}\label{SreptX}
	S^{rep,\tX}=
	\int d^4x \tLambda \tX \ .
\end{equation}
Then the second rearrangement of $\tX$ is performed in the following way
\cite{Kuchar:1991xd}
\begin{equation}
	\tX=\frac{1}{3!}\partial_t X^A \epsilon_{ABCD}X^B_iX^C_jX^D_k\epsilon^{ijk}
	\equiv \partial_tX^A n_A \ . 
\end{equation}
Using this form of $\tX$ we get from  the action (\ref{SreptX})  momenta conjugate to $X^A$ 
\begin{equation}\label{pAsecond}
p_A=\frac{\partial \mL}{\partial (\partial_t X^A)}=
\frac{1}{\kappa}\tLambda n_A \ . 
\end{equation}
Multiplying (\ref{pAsecond}) with $\partial_i X^A$ we obtain
\begin{equation}
p_A \partial_i X^A= 0 \ , \quad  i=1,2,3 \ 
\end{equation}
so that we have three constraints 
\begin{equation}
\mG_i\equiv p_A \partial_i X^A\approx 0 \ . 
\end{equation}
Let us explicitly determine form of $n_T$
\begin{equation}
n_T=\frac{1}{3!}\epsilon_{Tlmn}\partial_i Z^l\partial_j Z^m\partial_k Z\epsilon^{ijk}\equiv
\tilde{Z} \ , 
\end{equation}
where $\tilde{Z}$ is defined as determinant of the transformation $z\Rightarrow Z$
\begin{equation}
\det Z^m_i=\frac{1}{3!}\epsilon_{lmn}Z^l_iZ^m_jZ^n_k\epsilon^{ijk} \ . 
\end{equation}
Then we can express $\tLambda$ as
\begin{equation}
\tLambda=p_T \tilde{Z}^{-1} \ 
\end{equation}
and hence the  action has the form 
\begin{eqnarray}
S=\int d^4 x[\pi^{ij}\partial_t h_{ij}+p_\phi\partial_t \phi+p_A\partial_tX^A
-N(\mH_\tau+\frac{\tLambda}{\phi}\sqrt{h})-N^i\mH_i-v^i\mG_i] \ .
\nonumber \\
\end{eqnarray}
We see that the variation of the action with respect to $N,N^i$ and $v^i$ implies following primary constraints
\begin{eqnarray}
\mH^{rep,GR}\equiv \mH_\tau+\frac{\tLambda}{\phi}\sqrt{h}\approx 0 \ , \quad
\mH_i\approx 0 \ , \quad \mG_i\approx 0 \ . 
\nonumber \\
\end{eqnarray}
As we calculated in previous section the Poisson brackets between $\mH_\tau$ are closed and effectively decouple from the terms that depend on $p_A$ an $X^A$. 
The situation is different in case of the Poisson bracket between $\bT_T(\zeta)\equiv
\int d^3\bx \zeta(\mH_\tau+\frac{\tLambda}{\phi}\sqrt{h})$ and
$\bT_S(\xi^i)$ when 
\begin{eqnarray}\label{mHipre}
\pb{\bT_S(\xi^i),\bT_T(\zeta)}=\int d^3\bx \xi^m\partial_m \zeta (\mH_\tau
+\frac{\tLambda}{\phi}\sqrt{h})+\int d^3\bx \xi^m \zeta\frac{1}{\phi}\partial_m\tLambda
\nonumber \\
\end{eqnarray}
using
\begin{eqnarray}
	\pb{\bT_S(\xi^i),\mH_\tau}=
	-\xi^m
	\partial_m\mH_\tau-\partial_m\xi^m\mH_\tau \ . 
	\nonumber \\
\end{eqnarray}
The equation (\ref{mHipre}) implies that the constraint $\mH_i\approx 0$ is preserved during time evolution of
system if we impose the secondary constraint
\begin{equation}\label{secondcon}
\partial_m \tLambda \approx 0 \ . 
\end{equation}
We return to the solution of this constraint below. 
Let us introduce smeared form of the constraint $\mG_i$ in the form
\begin{equation}
\bG(\xi^m)=\int d^3\bx \xi^m\mG_m=\int d^3\bx \xi^m p_A\partial_m X^A
\end{equation}
so that we have
\begin{eqnarray}
&&\pb{\bG(\xi^m),p_A}=-\partial_m(\xi^m p_A) \ , \quad 
\pb{\bG(\xi^m),X^A}=-\xi^m\partial_m X^A  \ , \nonumber \\
&&\pb{\bG(\xi^m),Z_i^j}=
-\partial_i\xi^m Z_m^j-\xi^m\partial_m Z_i^j \ ,
\nonumber \\
&&\pb{\bG(\xi^m),\tilde{Z}}
=-\partial_m \xi^m\tilde{Z}-\xi^m\partial_m\tilde{Z} \ . 
\nonumber \\
\end{eqnarray}
Using these results we finally get
\begin{eqnarray}
\pb{\bG(\xi^m),\int d^3\bx \zeta
	 \frac{\tLambda}{\phi}\sqrt{h}}=
-\int d^3\bx \zeta\frac{\sqrt{h}}{\phi}\xi^m\partial_m \tLambda \approx 0
\nonumber \\
\end{eqnarray}
and hence it vanishes on the constraint surface $\partial_m\tLambda\approx 0$. In other words $\mG_i\approx 0 $ are first class constraints.

Let us return to the constraint (\ref{secondcon})
where we  deal with it in the following way. We write $\tLambda$ as
\begin{equation}
\tLambda=\hat{\tLambda}(\bx,t)+\tLambda_0 \ , \quad \int d^3\bx \sqrt{h}\hat{\tLambda}=0 \ , \quad 
\tLambda_0=\frac{1}{\int d^3\bx \sqrt{h}}\int d^3\bx \sqrt{h}\tLambda \ . 
\end{equation}
Then the constraint $\partial_m\tLambda=0$ implies that we have $3\times\infty^3-1$ constraints
$\hat{\tLambda}(\bx,t)\approx 0$ while $\tLambda_0(t)$ is not restricted. 

In the similar way we split $\mH_\tau$ as
\begin{equation}
	\mH_\tau(\bx,t)=\tilde{\mH}_\tau(\bx,t)+H_0\sqrt{h} \ , \quad 
	H_0=\frac{1}{\int d^3\bx \sqrt{h}}\int d^3\bx \mH_\tau
\end{equation}
so that
\begin{equation}\label{condmH}
	\int d^3\bx \tilde{\mH}_\tau=0 \ . 
\end{equation}
Then  the constraint $\mH^{rep,GR}\approx 0$ has the form 
\begin{eqnarray}
\mH_\tau+\frac{\tLambda}{\phi}\sqrt{h}=
\tilde{\mH}_\tau+H_0\sqrt{h}+\frac{\sqrt{h}}{\phi}\tLambda_0 \approx 0 \ . 
\nonumber \\
\end{eqnarray}
Integrating this equation over volume and using (\ref{condmH}) we obtain 
integral constraint
\begin{equation}\label{intcon}
H_0+\tLambda_0\frac{\int d^3\bx \frac{\sqrt{h}}{\phi}}{\int d^3\bx \sqrt{h}}
\approx 0  \ . 
\end{equation}

In summary we have the same constraint structure as in 
case of unimodular gravity where however the metric and dynamical variables are related
with the embedding variables through the global constraint 
(\ref{intcon}).

\section{Conclusion}\label{fourth}
In this work we studied gravity with dynamical determinant of metric in manifestly 
reparametrization invariant form where we used formalism introduced by K. Kucha\v{r} in 
\cite{Kuchar:1991xd}. This formalism is based on introducing four scalar fields that restore manifest diffeomorphism invariance of the action. Then  we shown that this theory reduces to specific form of unimodular gravity coupled to scalar field. In fact, the canonical structure of this model has the same form as in the case of unimodular gravity where the fact that the determinant of metric is dynamical does not bring new physical content to the theory except of the presence of new scalar field. It is remarkable that making theory manifestly diffeomorphism invariant solves all problems that were previously identified in theory with dynamical determinant of metric as for example the fact that the Poisson bracket between Hamiltonian constraint does not vanish on the constraint surface. We mean that this nicely demonstrate beauty and usefulness of the parametrization formulation of gravity formulated by K. Kucha\v{r}
in his papers \cite{Isham:1984rz,Isham:1984sb,Husain:1990vz}.

%

\appendix
\section{Calculation of Poisson bracket $\pb{\bT^{GR}_T(\xi),\bT_T^{GR}(\zeta)}$}
\setcounter{equation}{0}
In this appendix we present explicit calculations of the Poisson bracket
\begin{equation}
\pb{\bT_T^{GR}(\xi),\bT_T^{GR}(\zeta)}
\end{equation}
where the smeared forms of functions $\mH_\tau$ are defined as
\begin{equation}
\bT_T^{GR}(\xi)=\int d^3\bx \xi(\bx) \mH_\tau(\bx) \ , \quad 
\bT_T^{GR}(\zeta)=\int d^3\bx \zeta(\bx) \mH_\tau(\bx) \ ,
\end{equation}
where 
\begin{eqnarray}\label{mHApp}
&&\mH_\tau=
\frac{\kappa}{\sqrt{h}F}\Pi^{ij}\mM_{ijkl}
\Pi^{kl}-\frac{\kappa}{4GF\sqrt{h}}p_\phi^2
-\frac{1}{\kappa}F\sqrt{h}r-\nonumber \\
&&-\frac{1}{\kappa}FG\sqrt{h}h^{ij}\partial_i\phi\partial_j\phi+
\frac{2}{\kappa}\partial_i[F'\sqrt{h}h^{ij}\partial_j\phi] \ .
\nonumber \\
\end{eqnarray}
Note that we consider situation when  the matrix $\mM^{ijkl}$ is non-singular. 
In case of the singular matrix $\mM^{ijkl}$ we should replace $\mH_\tau$ given in  
(\ref{mHApp}) with $\mH_\tau^D$ defined in (\ref{repGRD}). Since the calculations and result are completely the same we will not write them explicitly here. 

We start with 
\begin{eqnarray}\label{PB1}
&&\pb{\int d^3\bx \xi\frac{\kappa}{\sqrt{h}F}	\Pi^{ij}\mM_{ijkl}\Pi^{kl},
	\int d^3\by \zeta \frac{1}{\kappa}F\sqrt{h}r}+\nonumber \\
&&\pb{
	\int d^3\bx \xi \frac{1}{\kappa}F\sqrt{h}r	,
	\int d^3\by \zeta\frac{\kappa}{\sqrt{h}F}	\Pi^{ij}\mM_{ijkl}\Pi^{kl}}=
\nonumber \\	
&&=-4\int d^3\bx (\zeta\nabla_m\xi-\xi\nabla_m\zeta)\Pi^{mn}\frac{\partial_nF}{F}+
\nonumber \\
&&+4\int d^3\bx (\zeta\nabla_m\xi-\xi\nabla_m\zeta)h^{mn}\frac{\partial_nF}{F}(1-2\bD)+\nonumber \\
&&+2\int d^3\bx (\zeta\partial_i\xi-\xi\partial_i\zeta)(\nabla_j\pi^{ij}-\frac{1}{2}h^{ij}\nabla_j(\frac{F'}{FG}p_\phi))\nonumber \\
&&+2\int d^3\bx (\zeta\partial_m\xi-\xi\partial_m\zeta)h^{mn}\nabla_n(\Pi(2\bD-1)) \ .  \nonumber \\
\end{eqnarray}
Further we calculate
\begin{eqnarray}
&&	\pb{\int d^3\bx \xi\frac{\kappa}{\sqrt{h}F}\Pi^{ij}\mM_{ijkl}\Pi^{kl},
	\frac{1}{\kappa}	\int d^3\by \zeta (-FG\sqrt{h}h^{ij}\partial_i\phi\partial_j\phi)}+\nonumber \\
&&+\pb{\frac{1}{\kappa}	\int d^3\bx \xi (-FG\sqrt{h}h^{ij}\partial_i\phi\partial_j\phi),
	\int d^3\by \zeta\frac{\kappa}{\sqrt{h}F}\Pi^{ij}\mM_{ijkl}\Pi^{kl}}=\nonumber \\	
&&=2\int d^3\bx  (\xi\partial_i\zeta-\zeta\partial_i\xi)\Pi (1-3\bD)h^{ij}\frac{\partial_jF}{F}
\nonumber \\
\end{eqnarray}
and
\begin{eqnarray}\label{PB3}
&&\pb{\int d^3\bx \xi \frac{\kappa}{\sqrt{h}F}\Pi^{ij}\mM_{ijkl}\Pi^{kl}, 
	\int d^3\by \zeta \frac{2}{\kappa}\partial_i[F'\sqrt{h}h^{ij}\partial_j\phi]}+\nonumber \\
&&\pb{\int d^3\bx \xi \frac{2}{\kappa}\partial_i[F'\sqrt{h}h^{ij}\partial_j\phi],
	\int d^3\by \zeta \frac{\kappa}{\sqrt{h}F}\Pi^{ij}\mM_{ijkl}\Pi^{kl}}=\nonumber \\
&&=-4\int d^3\bx (\xi\partial_m\zeta-\zeta\partial_m\xi)\Pi^{mn}
\frac{\partial_nF}{F}+\nonumber \\
&&+4\int d^3\bx (\xi\partial_m\zeta-\zeta\partial_m\xi)\Pi \bD h^{mn}\frac{\partial_nF}{F}+\nonumber \\
&&+2\int d^3\bx (\xi\partial_m \zeta-\zeta\partial_m\xi)\Pi (1-3\bD) h^{mn}\frac{\partial_nF}{F} -
\nonumber \\
&&-4\int d^3\bx(\xi\partial_m\zeta-\zeta\partial_m\xi)\frac{\Pi}{2}(1-3\bD)\frac{F'F''}{F^2G}h^{mn}
\partial_n\phi-\nonumber \\
&&-2\int d^3\bx (\xi\partial_m\zeta-\zeta\partial_m\xi)h^{mn}\sqrt{h}
\partial_n[\frac{1}{\sqrt{h}}\Pi(1-3\bD)\frac{F'}{F^2G}]F' \ . \nonumber \\
\end{eqnarray}
Further we have
\begin{eqnarray}
&&\pb{\int d^3\bx \xi(-\frac{\kappa}{4FG\sqrt{h}}p_\phi^2),
	\int d^3\by \zeta (-FG\sqrt{h}h^{ij}\partial_i\phi\partial_j\phi)}+
\nonumber \\
&&+	
\pb{
	\int d^3\bx \xi (-FG\sqrt{h}h^{ij}\partial_i\phi\partial_j\phi)	
	\int d^3\by \zeta(-\frac{\kappa}{4FG\sqrt{h}}p_\phi^2)}=
\nonumber \\
&&=\int d^3\bx (\xi \partial_m\zeta-\zeta\partial_m\xi)h^{mn}\partial_n\phi p_\phi
\nonumber \\	
\end{eqnarray}
and finally 
\begin{eqnarray}
&&\pb{-\int d^3\bx \frac{\xi\kappa}{4FG\sqrt{h}}p_\phi^2,
	2\int d^3\by \zeta\partial_m[F'\sqrt{h}h^{mn}\partial_n\phi]}+
\nonumber \\
&&\pb{
	\int d^3\bx 2\xi\partial_m[F'\sqrt{h}h^{mn}\partial_n\phi]	,
	-\int d^3\by \frac{\zeta\kappa}{4FG\sqrt{h}}p_\phi^2}=
\nonumber \\
&&=-\int d^3\bx (\xi\partial_m\zeta-\zeta\partial_m\xi)
h^{mn}\nabla_n(\frac{F'}{FG}p_\phi) \ . \nonumber \\
\end{eqnarray}
To proceed further we observe that we have
\begin{equation}
2\bD-1=\frac{\frac{F'^2}{F^2G}}{2-3\frac{F'^2}{F^2G}} \ , 
\quad 
1-3\bD
=-(2\bD-1)\frac{F^2G}{F'^2} \ . 
\end{equation}
Then collecting all terms together and using the fact that 
$\frac{\Pi}{\sqrt{h}}$ is scalar we obtain final  result
\begin{equation}
\pb{\bT^{GR}_T(\xi),\bT^{GR}_T(\zeta)}=
\bT_S((\xi\partial_m\zeta-\zeta\partial_m\xi)h^{mn}) \ . 
\end{equation}

{\bf Acknowledgement:}

This work  is supported by the grant “Dualitites and higher order derivatives” (GA23-06498S) from the Czech Science Foundation (GACR).

\end{document}